\documentclass[12pt]{article}

\usepackage{amsmath}
\usepackage{latexsym}
\newcommand\be{\begin{equation}}
\newcommand\ee{\end{equation}}
\newcommand\bea{\begin{eqnarray}}
\newcommand\eea{\end{eqnarray}}
\newcommand\beas{\begin{eqnarray*}}
\newcommand\eeas{\end{eqnarray*}}

\def\tr{{\rm Tr}}

\begin{document}
\title{Strongly coupled large N spectrum of two matrices coupled via a Yang-Mills interaction}

\author{\\
Martin N.~H. Cook and Jo\~ao P. Rodrigues\footnote{Email: joao.rodrigues@wits.ac.za} \\
\\
National Institute for Theoretical Physics, \\ 
School of Physics and Centre for Theoretical Physics, \\
University of the Witwatersrand, Johannesburg\\
Wits 2050, South Africa 
\\
}

\maketitle

\begin{abstract}
\end{abstract}
We consider the large $N$ spectrum of the quantum mechanical hamiltonian of two hermitean matrices 
coupled via a Yang-Mills interaction. In a framework where one of the matrices is treated 
exactly and the other is treated as a creation operator impurity, the difference equation associated with the Yang-Mills 
interaction is derived and solved exactly for two impurities. In this case, the full string tension corrected 
spectrum depends on two momenta. For a specific value of one of these momenta, the spectrum has the same structure as
that of giant magnon bound states. States with general number of impurities are also discussed.

\newpage

\noindent
\section{Introduction}

In this communication, we study the large $N$ spectrum of the quantum mechanical hamiltonian 

\be\label{Ham12}
\hat{H}
\equiv
\frac{1}{2} \tr (P_1^2 ) + \frac{w^2}{2}  \tr (X_1^2) + \frac{1}{2} \tr (P_2^2 )+ \frac{w^2}{2}  \tr (X_2^2)
- g_{YM}^2 \tr ([X_1,X_2][X_1,X_2]),
\ee

\noindent 
where $X_1$ and $X_2$ are two $N \times N$ hermitean matrices, and $P_1$ and $P_2$ their conjugate momenta, respectively. 


It is useful to think of (\ref{Ham12}) as associated with two of the six Higgs scalars of the bosonic sector of 
$\cal{N}=$ $4$ SYM, in the leading Kaluza Klein compactification on $S^3 \times R$. The harmonic potential results 
from the coupling to the curvature of the
manifold. Although we will only deal with bosonic degrees of freedom in this communication, the embedding of (\ref{Ham12}) 
in a supersymmetric setting is important as it allows one to neglect oscillator normal ordering contributions and to consider 
fluctuations about an unperturbed
single matrix background. In an angular momentum eigenstate basis (associated with $SO(2) \sim U(1) $ rotations in the $X_1-X_2$ plane), 
this single matrix background \cite{Corley:2001zk}, \cite{Berenstein:2004kk}  
has been shown to correspond 
\cite{Donos:2005vm}, \cite{Rodrigues:2005ec}, \cite{Rodrigues:2006gt} to a phase space description of the ``liquid'' drop model \cite{Lin:2004nb} 
of $ 1/2$  BPS states.

We will follow the approach first suggested in \cite{Donos:2005vm} of treating one of the matrices, $X_1$, exactly (in the large $N$ limit), and the other, $X_2$, 
referred to as an ``impurity" \cite{Berenstein:2002jq}, in a creation/annihilation basis. 
Letting

\be\label{Ide}
                             X_2 \equiv \frac{1}{\sqrt{2w}}( A_2 + A_2^{\dagger})  \quad P_2 = -i \sqrt{\frac{w}{2}}( A_2 - A_2^{\dagger}) 
\ee

\noindent
the hamiltonian (\ref{Ham12}) becomes, up to normal ordering terms,

\bea\label{HamMB}
\hat{H} &=& 
\frac{1}{2} \tr (P_1^2 ) + \frac{w^2}{2}  \tr (X_1^2) + w  \tr (A_2^{\dagger} A_2) \\
&-& \frac {g_{YM}^2}{2w} \tr ( 2 [X_1,A_2^{\dagger}][X_1,A_2] + [X_1,A_2]^2  + [X_1,A_2^{\dagger}]^2 ). \nonumber
\eea

\noindent
States can then be classified in terms of the number of $A_2$ impurities. 

The coupling of two matrices via a Yang-Mills interaction within the approach developed in \cite{Donos:2005vm} was extensively discussed in \cite{Rodrigues:2005ec}, 
where a ``square root" type hamiltonian was derived. Although two matrix systems expressed 
both in an angular momentum eigenstate basis and real-imaginary basis were discussed, only the first order (in $g_{YM}^2N$) correction 
to the energies was analyzed. 
In this communication, we will not complexify the Higgs, and will treat them as two hermitean (``real") matrices, but we will carry out a full study of the spectrum.

This paper is organized as follows: in Section $2$, a description of the method used to obtain the relevant large $N$ background and fluctuations is provided.
To ensure its validity, a Bogoliubov transformation, as first suggested in \cite{Rodrigues:2005ec}, is introduced resulting in a square root type hamiltonian.
In Section $3$, the spectrum of fluctuations of invariant states and corresponding eigenstates about the free harmonic background system 
\cite{Rodrigues:2005ec} is reviewed. In Section $4$, the effect of the Yang-Mills interaction on two impurity states is studied, and an exact spectrum is obtained. First, the
mixing of states resulting from the action of the term linear in $g_{YM}^2N$ is described in terms of a {\it{two}} index difference equation. This difference 
equation is then diagonalized and summed. For a specific value of one of the conjugate momenta, this spectrum agrees with the bound state spectrum of two magnons  
\cite{Hofman:2006xt}. In Section $5$, states with an arbitrary number of impurities are considered and their first order (in $g_{YM}^2N$) spectrum obtained. 
Conclusions and a brief discussion are included.

\noindent
\section{Method and square root hamiltonian}

\noindent
We are interested in the background and spectrum of gauge invariant states. One way to implement this invariance in the large $N$
limit, is to restrict the hamiltonian to act on wave functionals of invariant single trace operators (``loops").

A complete set (in the large $N$ limit) of gauge invariant operators is given by \footnote{In this article, we will often
switch from creation-annihilation operators to their coherent state representation $A_2^{\dagger}\to A_2$,
$A_2\to\partial/{\partial A_2}$.}    :

\bea
\psi(k;0)  &=&  \tr(e^{ikX_1})   \nonumber\\
\psi(k;1)  &=&  \tr(e^{ikX_1} A_2 )\nonumber  \\
\psi(k_1,k_2;2)  &=&  \tr(e^{ik_1 X_1} A_2 e^{ik_2 X_1} A_2)\\
           & ... \nonumber \\
\psi(k_1,k_2,...,k_s;s)  &=&  \tr \big( \prod_{i=1}^s (e^{ik_i X_1} A_2 )\big), \quad s \rangle 0. \nonumber 
\eea

Equivalently,

\bea\label{Densities}
\psi(x;0)&=& \int \frac{dk}{2\pi} e^{-ikx} \psi(k;0)  =  \tr(\delta(x-X_1))   \nonumber\\
\psi(x;1)&=& \int \frac{dk}{2\pi} e^{-ikx} \psi(k;1)  =  \tr(\delta(x-X_1)A_2)    \\
\psi(x_1,x_1;2)&=& \int \int \frac{dk_1}{2\pi}\frac{dk_2}{2\pi} e^{-ik_1x_1} e^{-ik_2 x_2} \psi(k_1,k_2;2) \nonumber \\
&=&  \tr(\delta(x_1-X_1) A_2 \delta(x_2-X_1) A_2 )\nonumber \\
           & ... \nonumber \\
\psi(x_1,x_2,...,x_s;s)&=& \int ...\int \frac{dk_1}{2\pi}... \frac{dk_s}{2\pi}  
                        e^{-ik_1x_1} ... e^{-ik_sx_s}  \psi(k_1,k_2,...,k_s;s)\nonumber  \\
&=& \tr \big( \prod_{i=1}^s (\delta(x_i -X_1) A_2 )\big), \quad s \rangle 0. \nonumber 
\eea

\noindent
We will refer to these as ``$s$ impurity states" and, to simplify notation, we will often denote them 
by $\psi(A;s)$, with $A$ an appropriate generic index. 

The restriction of the action of the hamiltonian (\ref{HamMB}) on functionals of the invariant operators is
implemented by performing a change of variables \cite{Jevicki:1979mb} from the original matrix variables to
the invariant variables. Because of the reduction in the number of degrees of freeedom, 
the Jacobian of this transformation has to be taken into account \cite{Jevicki:1979mb}. 

If one can ensure that loops with non-vanishing number of impurities have vanishing expectation values, i.e.,

\be\label{Vanish}
\langle  \psi(k_1,k_2,...,k_s;s) \rangle = \langle e^{ik_1 X_1} A_2 e^{ik_1 X_1} A_2 ... e^{ik_s X_1} A_2 \rangle = 0 \quad s\ne 0 ,
\ee

\noindent
this Jacobian only depends on the zero impurity variables \cite{Donos:2005vm}. 
Therefore, only the usual large $N$ single matrix background of the matrix $X_1$ is generated.

In order to ensure that (\ref{Vanish}) is verified, we perform a Bogoliubov transformation

$$
     {A_2}_{ij} = \cosh (\phi_{ij}) B_{ij} - \sinh (\phi_{ij}) B^{\dagger}_{ij}  
$$ 

\noindent
with

$$
 \tanh(2\phi_{ij}) = \frac{\frac {g_{YM}^2}{w}(\lambda_i-\lambda_j)^2}{w + \frac{g_{YM}^2}{w}(\lambda_i-\lambda_j)^2}
$$

\noindent
where the $\lambda_i$'s are the eigenvalues of the matrix $X_1$. Then (\ref{HamMB}) takes the form

\be\label{Htot}
\hat{H} = \frac{1}{2} \tr (P_1^2 ) + \frac{w^2}{2}  \tr (X_1^2) +
\sum_{i,j=1}^N \sqrt{w^2 + 2 {g_{YM}^2}(\lambda_i-\lambda_j)^2} 
\quad\bar{B}^{\dagger}_{ij}\bar{B}_{ji},  
\ee

\noindent
where $\bar{B}=V^{\dagger} B V$, with $V$ the unitary matrix that diagonalizes $X_1$. Normal ordering contributions
are again not included.

\subsection{Background and fluctuations about the $g_{YM}=0$ theory}

The $X_1$ background and zero impurity spectrum of fluctuations are associated with the large $N$ hamiltonian dynamics of a single hermitean matrix.  
This is completely described by the standard cubic collective field hamiltonian 
(in addition to the original derivation \cite{Jevicki:1979mb}
and its aplication to $c=1$ strings \cite{DasKA},\cite{Demeterfi:1991tz}, reference \cite{deMelloKoch:2002nq}
also has a general self-contained review of the method).    

Including only the terms required for a study of the background and fluctuations,
this hamiltonian takes the form:

\be\label{CollBef}
{- \frac{1}{2}}
\int dx \partial_x {\partial \over \partial \psi(x,0)} \psi(x,0) \partial_x {\partial \over \partial \psi(x,0)}
+ \int dx \Big( {\pi^2 \over 6}\psi^3(x,0) +  \psi(x,0)({w^2 x^2 \over 2}- \mu) \Big)
\ee

\noindent
where the Lagrange multiplier $\mu$ enforces the contraint

\be \label{Const}
\int dx \psi(x,0) = N .
\ee

\noindent
To exhibit explicitly the $N$ dependence, we rescale

\bea \label{Rescaling}
x  \to \sqrt{N} x  &\quad&  \psi(x,0) \to \sqrt{N} \psi(x,0)   \nonumber   \\
-i {\partial \over \partial \psi(x,0)}\equiv \Pi(x)  \to {1 \over N} \Pi(x) &\quad&  \mu \to N \mu\ 
\eea

\noindent
and obtain

\bea \label{HEffZero}
H_{eff}^{0}&=& {{1 \over 2N^2}}
\int dx \partial_x \Pi(x) \psi(x,0) \partial_x \Pi(x) \\ 
&+& N^2 \Big(  \int dx {\pi^2 \over 6}\psi^3(x,0) +  \psi(x,0)({w^2 x^2 \over 2}- \mu) \Big), \nonumber
\eea

\noindent
giving the well known Wigner distribution background in the limit as $N \to \infty$:

\be\label{po}
\pi \psi(x,0) \equiv \pi \phi_0 = \sqrt{2\mu -w^2 x^2} = \sqrt {2 w - w^2 x^2}.
\ee

\noindent
For the zero impurity small fluctuation spectrum, one shifts 

$$
\psi(x,0) = \phi_0 + {1\over \sqrt{\pi} N} {\partial_x \eta };
\qquad \partial_x \Pi(x) = - \sqrt{\pi} N P (x)
$$

\noindent
to find the quadratic operator

$$
H_{2}^{0}= {{1 \over 2}}
\int dx (\pi\phi_{0}) P^2(x) + {1 \over 2} \int dx (\pi\phi_0) ({\partial_x \eta })^2
$$

\noindent
By changing to the classical ``time of flight" $\phi$ {\footnote{In an angular momentum eigenstate basis, $\phi$ has a clear gravity interpretaton \cite{Donos:2005vm}, 
as the angular variable in the plane of the droplet \cite{Lin:2004nb}}}:

\be\label{phio}
{dx \over d\phi} = \pi \phi_0 ; \quad x(\phi)= - \sqrt{\frac{2}{w}} \cos(w\phi) ; \quad \pi \phi_0 =  \sqrt{2 w} \sin (w \phi),
\quad 0 \le \phi \le \frac{\pi}{w} ,
\ee

\noindent
one obtains the hamiltonian of a free $1+1$ massless boson \cite{DasKA}:

\be\label{Quadratic}
H_{2}^{0}= {{1 \over 2}}
\int d\phi P^2(\phi) + {1 \over 2} \int d\phi ({\partial_\phi \eta })^2
\ee

Imposing Dirichelet boundary conditions at the classical turning points yields the spectrum in the zero impurity sector 

\be \label{SpecZero}
                  \epsilon_j= w j  \quad ; \qquad     \phi_j = \sin(j w \phi),  \quad j=1,2 , ...
\ee

\noindent
In \cite{Donos:2005vm} it was shown that the form of the quadratic operator determining the many-impurity spectrum is

\be \label{HTwoMulti}
H_2^{s} = {-\frac{1}{2}} \sum_{A} \bar{\omega} (A,s) {\partial \over \partial \psi(A,s)}
+ {1\over 2} \int dx \sum_{A} \Omega(x,0:A,s)
          {\partial \ln J \over \partial \psi(x,0)} {\partial \over \partial \psi(A,s)}
\ee

\noindent
where, to leading order in $N$

\be \label{Jac}
\partial_x {\partial \ln J \over \partial \psi(x,0)} = \partial_x \int dy \Omega^{-1}(x,0;y,0) \omega(y,0) =
2 \int dy {\phi_0(y) \over (x-y)}
\ee

\noindent
In (\ref{HTwoMulti}) and (\ref{Jac}), $\omega (A,s)$ and $\Omega(x,0:A,s)$ have their usual meanings \cite{Jevicki:1979mb}:

\bea \label{LapK}
\omega (A,s)&=& \tr \left({\partial^2 \psi(A,s) \over \partial M^2}\right) \\
\Omega (x,0:A,s)&=& \tr \left({\partial\psi(0,x) \over \partial M}{\partial\psi(A,s) \over \partial M} \right).
\eea

\noindent
$\omega (A,s)$ splits the loop $\psi(A,s)$ and $\Omega (A,s:A',s')$ joins the two loops $\psi(A,s)$ and $\psi(A',s')$.
$\bar{\omega}(A,s)$ indicates that only splittings into zero impurity loops need be considered.

A detailed analysis of the eigenvalues and eigenfunctions of (\ref{HTwoMulti}) for the set of multi-impurity states (\ref{Densities}) in the harmonic background 
has been carried out in \cite{Rodrigues:2005ec}. The result is that the eigenstates and eigenvalues with $s$ impurities are

\bea\label{freenergies}
\epsilon_{j_i}= w \Big(\sum_{i=1}^s j_i \Big) &;& 
\langle \psi | j_1,j_2,...,j_s \rangle = \tr({u_{j_1}(X_1)} B {u_{j_2}(X_1)} B ... {u_{j_s}(X_1)}B) \nonumber \\
 &;& j_i=0,1,2,...,
\eea

\noindent
where the polynomials $u_{j}(x)$ are rescaled Chebyshev polynomials of the second kind $U_j(x)$:

\be\label{uu}
u_{j}(x) \equiv  {\sin ((j+1) w \phi(x)) \over \sqrt{2 w} \sin(w \phi(x))}=  {\sin ( (j+1) w \phi(x)) \over \pi\phi_0(x)} =
 \frac{1}{\sqrt{2w}} U_{j}\left(-\sqrt{\frac{w}{2}} x\right).
\ee

\noindent
These states are in one to one correspondence with the spectrum of the states
$\tr({A_1^{\dagger}}^{j_1} {B^{\dagger}}{A_1^{\dagger}}^{j_2}{B^{\dagger}}...{B^{\dagger}})$, 
with
(${\sqrt{2w}} X_1 \equiv ( A_1 + A_1^{\dagger})$),
obtained by acting on the
original Fock space of the theory.

\section {$g_{YM}$ interactions - two impurity spectrum}

We now return to the hamiltonian (\ref{Htot}), after rescaling according to (\ref{Rescaling}). We wish to study the interaction part of this hamiltonian:

\bea\label{HYM}
\hat{H}_{YM} &\equiv& \hat{H} - \left( \frac{1}{2N} \tr (P_1^2 ) + \frac{w^2N}{2}  \tr (X_1^2) \right) \nonumber \\
&=& \sum_{i,j=1}^N \sqrt{w^2 + 2 {g_{YM}^2N}(\lambda_i-\lambda_j)^2} 
\quad\bar{B}^{\dagger}_{ij}\bar{B}_{ji} \\
&=& \sum_{n=0}^{\infty} A_n w \left( \frac{2 {g_{YM}^2N}}{w^2}\right)^n  \hat{O}_n \nonumber
\eea  
 
\noindent
with $A_n = {1/2 \choose n}$ and 

\bea
\hat{O}_n &\equiv& (-1)^n \tr ([X_1,[\underbrace{X_1,...[X_1,B^{\dagger}]...}]][X_1,[\underbrace{X_1,...[X_1,B]...}]]). \nonumber \\
&&\qquad \qquad\qquad\qquad\quad ~  \textrm{\textrm{$n$ nested commutators}} \nonumber
\eea

We will first consider the action of the first non trivial term in the expansion on a two impurity state:

\bea
\hat{O}_1 \tr({u_{j_1}(X_1)} B^{\dagger} {u_{j_2}(X_1)} B^{\dagger}) &=& 
- \tr ([X_1,B^{\dagger}][X_1,B]) \tr({u_{j_1}(X_1)} B^{\dagger} {u_{j_2}(X_1)} B^{\dagger}) \nonumber \\ 
&\equiv& \hat{O}_1 | j_1,j_2 \rangle
\eea

\noindent
We obtain \cite{Rodrigues:2005ec}:

\bea
\hat{O}_1 \tr({u_{j_1}(X_1)} B^{\dagger} {u_{j_2}(X_1)} B^{\dagger})
&=& - 4 \tr(X_1 {u_{j_1}(X_1)} B^{\dagger} X_1 {u_{j_2}(X_1)} B^{\dagger}) \nonumber \\
&+& 2 \tr(X_1^2 {u_{j_1}(X_1)} B^{\dagger} {u_{j_2}(X_1)} B^{\dagger}) \\
&+& 2 \tr({u_{j_1}(X_1)} B^{\dagger} X_1^2 {u_{j_2}(X_1)} B^{\dagger}) \nonumber
\eea

\noindent
The right hand side can be calculated using the identity

$$
U_{j+1}(x)-2 x U_{j}(x) + U_{j-1}(x)=0.
$$

\noindent
In \cite{Rodrigues:2005ec} only terms contributing to first order perturbation theory (i.e., preserving $j_1+j_2$) were considered. The full result is:

\bea\label{differencetwo}
w \hat{O}_1 | j_1,j_2 \rangle &=& 4 | j_1,j_2 \rangle - 2 | j_1+1,j_2-1 \rangle - 2 | j_1-1,j_2+1 \rangle  \nonumber \\
                                          &-& 2 | j_1+1 ,j_2+1 \rangle + | j_1+2,j_2 \rangle + | j_1,j_2+2 \rangle \\
                                          &-& 2 | j_1-1 ,j_2-1 \rangle + | j_1-2,j_2 \rangle + | j_1,j_2-2 \rangle \nonumber
\eea

\noindent
In terms of $j=j_1, J=j_1+j_2$ (these can be thought of as the ``positions" of the two impurities\footnote{In our notation, the last impurity is always placed at
the end of the ``string", so $J$ is also the length of the string}), these recurrence relations take the form:

\bea\label{differenceone}
w \hat{O}_1 |j,J \rangle &=& 4 |j,J \rangle - 2 |j+1,J \rangle - 2 |j-1,J \rangle \nonumber \\
                                          &-& 2 | j+1 ,J+2 \rangle + | j+2,J+2 \rangle + | j,J+2 \rangle \\
                                          &-& 2 | j-1 ,J-2 \rangle + | j-2,J-2 \rangle + | j,J-2 \rangle \nonumber
\eea

\noindent
We remark that these difference equations generalize those so far considered in the literature (\cite{Staudacher:2004tk} and references therein).
In addition to the standard $J$ conserving terms, we observe the presence of $J$ non-conserving terms. This is not surprising, given that our matrices
do not carry $U(1)$ charges. Diagrams responsible for transitions $ J \to J \pm 2 $ can be easily identified \cite{Martin}.

In this communication, we study the eigenvalues of the difference operator (\ref{differencetwo}), i.e., we consider generic $j_1$ and $j_2$ values. Letting

$$
| j_1,j_2 \rangle \sim \lambda_1^{j_1} \lambda_2^{j_2} 
$$

\noindent
We obtain

$$
w \hat{O}_1  \lambda_1^{j_1} \lambda_2^{j_2} = \Big( 
     4 - 2 \frac{\lambda_1}{\lambda_2} - 2 \frac{\lambda_2}{\lambda_1} - 2 \lambda_1 \lambda_2 + \lambda_1^2 + \lambda_2^2 
                       - 2 \frac{1}{\lambda_1 \lambda_2} + \frac{1}{\lambda_1^2} + \frac{1}{\lambda_2^2}
\Big) \lambda_1^{j_1} \lambda_2^{j_2}
$$

\noindent
Hermiticity is satisfied if a plane wave basis is considered:

$$
    \lambda_1 = e^{i k_1} \quad \lambda_2 = e^{i k_2}.
$$

\noindent
The energies $E^1_{(k_1,k_2)}$ of the operator

\be\label{Hone}
\hat{H}_1 \equiv \frac{g_{YM}^2 N}{w}  \hat{O}_1
\ee

\noindent
are then

$$
E^1_{(k_1,k_2)} = \frac{16 g_{YM}^2 N}{w^2} \sin^2\left(\frac{k_1-k_2}{2} \right) \sin^2\left(\frac{k_1+k_2}{2}\right)
$$

In terms of momenta $p_1$ and $p_2$ conjugate to the positions of the impurities ($p_1=k_1-k_2, p_2=k_2$),

\be\label{twospect}
E^1_{(p_1,p_2)} = \frac{16 g_{YM}^2 N}{w^2} \sin^2\left(\frac{p_1}{2}\right) \sin^2\left(p_2 + \frac{p_1}{2}\right)
\ee

It turns out that it is possible for two impurity states to calculate the full contribution of the square root in (\ref{HYM}). This is a result of
the property: 

\be\label{prop}
\hat{O}_n   | j_1,j_2 \rangle = \frac{1}{2^{n-1}} \hat{O}_1^n | j_1,j_2 \rangle.
\ee

This property can be proved by induction \cite{Martin}. As a result, we obtain for the exact energy of the two impurity state

$$
E_{(p_1,p_2)} = 2 \sqrt{w^2 + \frac{16 g_{YM}^2 N}{w} \sin^2\left(\frac{p_1}{2}\right) \sin^2\left(p_2 + \frac{p_1}{2}\right)}
$$

The spectrum of a system is one of its most important characteristics, so it may be possible that it remains unchanged even when using states of definite 
angular momentum. An indication that this may be true is to consider the first order (in $\lambda= g_{YM}^2 N)$ result (\ref{twospect}) when 
$p_2= \frac{\pi}{2}$. One obtains ($w=1$)

$$
2 + \frac{16 g_{YM}^2 N}{w^2} \sin^2\left(\frac{p_1}{2}\right) \cos^2\left(\frac{p_1}{2}\right)
$$

This result has the same structure as that of the perturbative spectrum of the two giant magnon bound state obtained in \cite{Hofman:2006xt}.

\section{More impurities and conclusion}

The action of the operator $\hat{O}_1$ on a state with $s$ impurities yields the following difference equations:

\bea\label{differencetmany}
w \hat{O}_1 | j_1,j_2,...,j_s \rangle &=& 2 | j_1,j_2,..,j_s \rangle \nonumber \\
                                          &-& | j_1+1,j_2-1,..,j_s \rangle - | j_1-1,j_2+1,,..,j_s \rangle  \nonumber \\
                                          &-& | j_1+1 ,j_2+1,,..,j_s \rangle + | j_1+2,j_2,,..,j_s \rangle \\
                                          &-& | j_1-1 ,j_2-1,,..,j_s \rangle + | j_1-2,j_2,..,j_s \rangle\nonumber\\
                                          &+& \text{cyclic permutations} \nonumber
\eea

This difference operator is again diagonalized by plane wave states

$$
 | j_1,j_2, ... , j_s \rangle \sim  e^{i k_1 j_1}  e^{i k_2 j_2}  ... e^{i k_s j_s}   ,                            
$$

with the result that the eigen-energies of the difference operator $H_1$ in (\ref{Hone}) are:

$$
E^1_{(k_1,k_2,...,k_s)} = \frac{4 g_{YM}^2 N}{w^2} \sum_{i=1}^{s} \big(  \cos^2(k_i) - \cos^2(k_i) \cos^2(k_{i+1}) \big) \quad k_{s+1}\equiv k_{1}
$$
 
However, for more than $2$ impurities, property (\ref{prop}) or equivalent no longer holds\footnote{When acting on three impurity states, it turns that one still has 
$\hat{O}_1^2 = 2 \hat{O}_2$.}.

In summary, by expanding about an asymmetric large $N$ background where only one of two hermitiean matrices acquires a large $N$ expectation value, and perturbing 
about this configuration with creation/annhilition impurities of the other matrix in the presence of a Yang-Mills interaction, 
the full string tension corrected spectrum of two impurity states and the leading (in $g_{YM}^2N$) 
spectrum of multi-impurity states has been obtained.       
For two impurities, and for special values of momenta, the perturbative spectrum of the two giant magnon bound state obtained 
in \cite{Hofman:2006xt} is recovered\footnote{A matrix derivation of this result has also been obtained \cite{Berenstein:2007zf}
in the context of an approach based on mutually commuting matrices \cite{Berenstein:2005jq}. See also \cite{Hatsuda:2006ty}.}. 

\section{Acknowledgements}
One of us (J.P.R.) would like to thank the hospitality of the High Energy Theory Group at Brown University during 
his visit, when this communication was (finally!) written up. This research is supported by a 
NRF Focus Areas Grant (Grant Unique Number 2053791).

\end{document}